\pdfoutput = 1
\documentclass[journal=nalefd, manuscript=letter]{achemso}
\usepackage{graphicx}
\usepackage{graphics}
\usepackage{amsmath}
\usepackage{amssymb}
\usepackage{amsfonts}
\usepackage{dcolumn}
\usepackage{dsfont}
\usepackage{latexsym}
\usepackage{rotating}
\usepackage{array}
\usepackage{makecell}
\usepackage{color}
\usepackage{latexsym}
\usepackage{bbm}
\usepackage{physics}
\usepackage{subfigure}
\usepackage{float}
\usepackage{epsfig}
\usepackage{epsf}
\usepackage{psfrag}
\usepackage{bm}
\usepackage{amsthm}
\usepackage{eucal}
\usepackage{mathrsfs}
\usepackage{tabularx}
\usepackage{url}
\usepackage{braket}
\usepackage{epstopdf}
\usepackage{color} 

\usepackage{hyperref}
\hypersetup{
colorlinks=true,
        final=true,
        linkcolor=black,
        citecolor=black,
        filecolor=black,
        urlcolor=black,
}

\title{Octupolar vortex crystal and toroidal moment in twisted bilayer MnPSe$_3$}

\author{Muhammad Akram}
\affiliation{Department of Physics, Arizona State University, Tempe, AZ 85287, USA}
\affiliation{Department of Physics, Balochistan University of Information Technology, Engineering and Management Sciences (BUITEMS), Quetta 87300, Pakistan}
\author{Fan Yang}
\affiliation{Department of Chemical Engineering and Materials Science, University of Minnesota, Minneapolis, MN 55455, USA}
\author{Turan Birol}
\affiliation{Department of Chemical Engineering and Materials Science, University of Minnesota, Minneapolis, MN 55455, USA}
\author{Onur Erten}
\affiliation{Department of Physics, Arizona State University, Tempe, AZ 85287, USA}
\email{onur.erten@asu.edu}

\begin{document}


\begin{abstract}
{Experimental detection of antiferromagnetic order in two-dimensional materials is a challenging task due to the absence of net dipole moments. Identifying multi-domain antiferromagnetic textures via the current techniques is even more difficult. In order to address this challenge, we investigate the higher order multipole moments in twisted bilayer MnPSe$_3$. While the monolayers of MnPSe$_3$ exhibit in-plane N\'eel antiferromagnetic order, our atomistic simulations indicate that the moir\'e superlattices display a two-domain phase on each layer. We show that the octupolar moments $M_{33}^+$ and $M_{33}^-$ are significant in this multi-domain phase at the domain walls. In addition, when  $[M_{33}^+,M_{33}^-]$ are represented by the $x$ and $y$ components of a vector, the resultant pattern of these octupole moments winds around the antiferromagnetic domains and forms to vortex crystals which leads to octupolar toroidal moments, $T_{xyz}$ and $T_{z}^{\beta}$. $T_{xyz}$ and $T_{z}^{\beta}$ can give rise to a magnetoelectric effect and gyrotropic birefringence that may provide indirect ways of detecting multi-domain antiferromagnetic order. Our results highlight the importance of higher-order multipole moments for identification of complex spin textures in moir\'e magnets.}

\begin{center}
\end{center}

\end{abstract}

Two-dimensional (2D) magnets are intriguing materials that provide distinct possibilities for both fundamental research and technological advancements.\cite{Huang2017Nature,Gong2017Nature,Fei2018NatureMaterials,Cai2019ACS,Long2020Nano_Letters,Kang2020NatureCommunications, Blei_APR2021}.  
Among 2D vdW magnets, MPX$_3$ (M = Mn, Fe, Ni, Co; X= S, Se) represents a class of antiferromagnetic (AFM) materials with relatively high transition temperatures displaying both N\'eel and zigzag AFM order\cite{Joy_PRB1992}. In these materials, transition metal ions carry localized magnetic moments arranged in a layered honeycomb lattice structure\cite{OUVRARD19851181,Du2016ACSNano,Wang2018AFM}. Their magnetic properties are sensitive to both transition metal and chalcogen ions and can be tuned by the choice of M and X\cite{Wang2018AFM,Joy1992PRB,Wildes2015PRB,Rule2007PRB,PhysRevB.94.214407,Rabindra2021PRM,Shemerliuk2021EM,Afanasiev2020,Rabindra2022PRM}. Within this class, MnPSe$_3$ is a N\'eel AFM with in-plane magnetization and T$_N = 74$K\cite{Calder2021PRB,Rabindra2022PRM}. Detecting AFM ordering in 2D magnets is a challenge due to zero net dipole moment. While Raman spectroscopy is one of the effective techniques for detecting long range AFM ordering in 2D magnets\cite{Kim_2019},  detecting multi-domain structures in 2D AFMs is even more challenging. Hence, higher-order multipole moments in multi-domain structures could provide alternative means for detecting complex spin textures.

Moir\'e superlattices arising from twisted bilayers of van der Waals (vdW) materials represent an ideal platform for studying a wide range of novel phenomena including unconventional superconductivity, fractional Chern insulators and Wigner crystals, to name a few\cite{Cao2018, Xie_Nature2021, Regan_Nature2020}. 
Although there has been a thorough research on moir\'e engineering of electronic phases, the research efforts on moir\'e superlattices consisting of magnetic materials is in their early stages. Several novel non-coplanar phases have been predicted in moir\'e magnets\cite{Tong_ACS2018, Hejazi_PNAS2020, Hejazi_PRB2021, Akram_PRB2021, Akram_NanoLett2021, Tong_PRR2021, Xiao_PRR2021, Ghader_CommPhys2022, Zheng_AFM2023, Kim_NanoLett2023, Fumega_2DMAtt2023, Keskiner_NanoLett2024, Akram_NanoLett2024, Das_arXiv2024, Antao_arXiv2024, Kim_arXiv2024} and recent experiments have confirmed some of these phases\cite{Xu2022NatureNano, Song_Science2021, Hongchao2023NaturePhysics, Huang_NatComm2023, Cheng_NatElect2023, Sun_NatPhys2024}. Since monolayers of MPX$_3$ can be isolated by exfoliation methods\cite{Wang2018AFM,Wildes2015PRB,Rule2007PRB,PhysRevB.94.214407,Rabindra2022PRM} and heterostructures with other vdW materials such as transition-metal dichalcogenides can be constructed\cite{Onga2020ACSNano},  moir\'e engineering in MPX$_3$ can be explored as a method to control its magnetism and create novel magnetic textures\cite{Sun_NatPhys2024}.

Here, we investigate the magnetic phases and the multipolar order in twisted bilayer MnPSe$_3$ using atomistic simulations. Our main findings are summarized as follows: (i) we obtain a two-domain phase in each layer, where the angle between the order parameters on each domain increases with increasing interlayer coupling and decreasing twist angle, until it reaches 90$^\circ$. (ii) We show that the dipole moments at the domain walls are negligible. Among different types of higher order multipolar order parameters, we determine that two types of octupolar moments, $M_{33}^+,M_{33}^-$ to be significant. (iii) A two-component vector consisting of $[M_{33}^+,M_{33}^-]$ forms vortex crystals, with one vortex per moir\'e unit cell at each layer. (iv) These octupolar vortices give rise to finite octupolar toroidal moment which results in magnetoelectric effect and gyrotropic birefringence.  

\begin{figure}
\center
\includegraphics[width=0.9\textwidth]{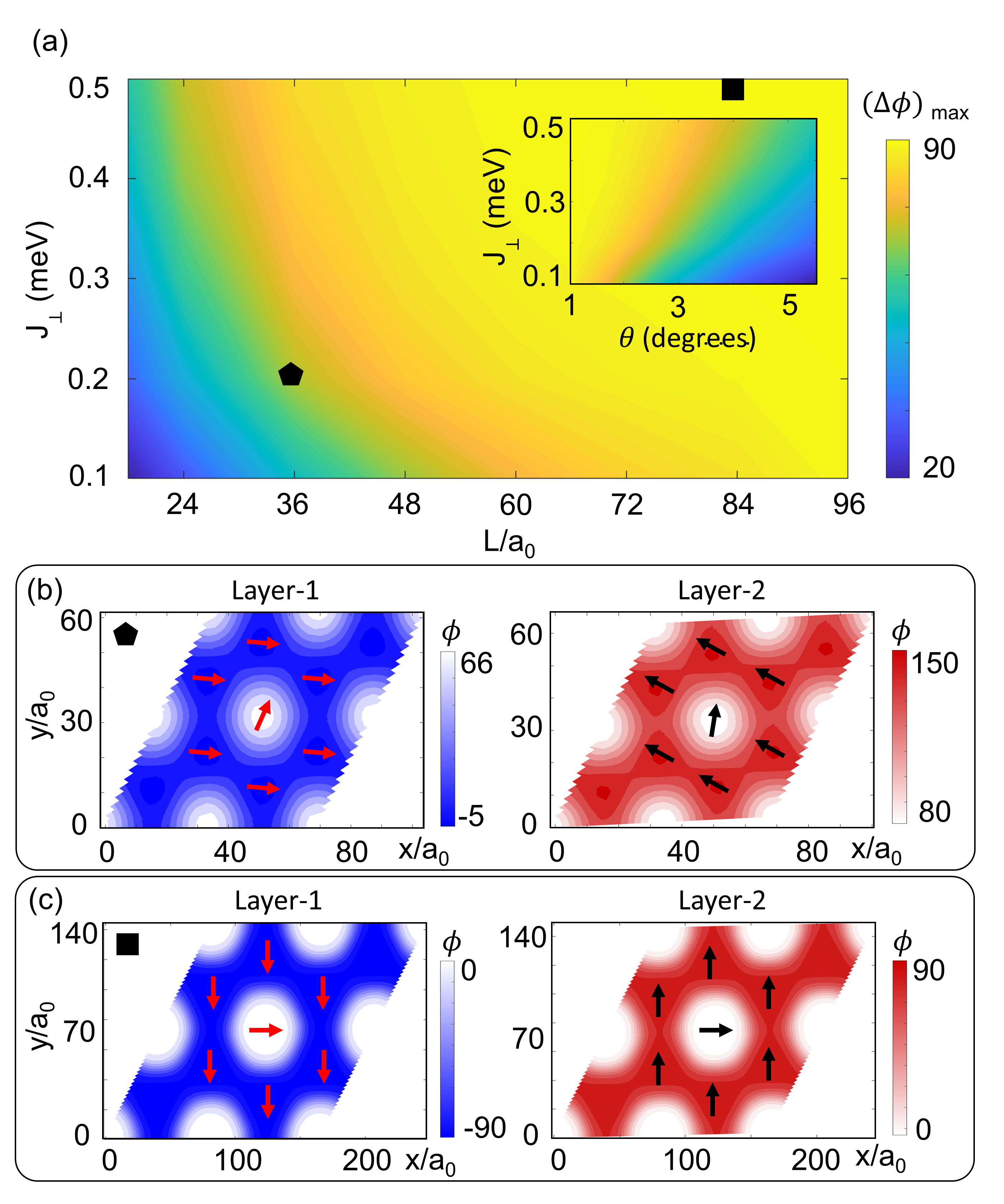} 
\caption{(a) Phase diagram of twisted bilayer MnPSe$_3$ as a function moir\'e period $L$ and interlayer coupling J$_{\perp}$. The color represents the maximum change in the relative orientation of order parameter between AA stacking and AB stacking regions on each layer. The inset shows the same phase diagram as a function twist angle $\theta$ and interlayer coupling J$_{\perp}$. (b) Orientation of the order parameter in the two layers at $L=36a_0$ and J$_{\perp}$=0.2 meV for $2\times2$ moir\'e unit cells. The arrows represent the direction of order parameter at AA and AB stacking. (c) Orientation of order parameter in the two layers at $L=84a_0$ and J$_{\perp}$=0.5 meV for $2\times2$ moir\'e unit cells.}
\label{Fig:1}
\end{figure}
We start by introducing the spin Hamiltonian which describes the magnetic properties of twisted bilayer  MnPSe$_3$:
$\mathcal{H} = \mathcal{H}_{intra}^1+\mathcal{H}_{intra}^2+\mathcal{H}_{inter}$
where $\mathcal{H}_{intra}^{1 (2)}$ contains the intralayer exchange terms in layer 1 (2) and $\mathcal{H}_{inter}$ contains the interlayer exchange, 
\begin{eqnarray} 
\label{Eq:Gnrl_Hmlt_1}
\mathcal{H}_{intra}&=&J_1\sum_{\langle ij \rangle}  \mathbf{S}_{i}\cdot \mathbf{S}_{j} +J_2 \sum_{\langle \langle ij \rangle \rangle} \mathbf{S}_{i}\cdot \mathbf{S}_{j}+J_3 \sum_{\langle \langle \langle ij \rangle \rangle \rangle} \mathbf{S}_{i}\cdot \mathbf{S}_{j} +A\sum_{i} ({S}_{i}^z)^2,\\
\mathcal{H}_{inter}&=& \sum_{\langle ij \rangle} J_\perp({ r}_{ij}) {\bf S}_i^{1}\cdot {\bf S}_j^{2}.
\end{eqnarray}
Here $J_1$, $J_2$ and $J_3$ are the first, second and third nearest-neighbor Heisenberg exchange couplings and $A$ is the single ion anisotropy. $J_\perp({r}_{ij})$ is the interlayer exchange coupling and ${ r}_{ij}$ represents the interlayer displacement. 
For MnPSe$_3$ the intralayer exchange parameters are antiferromagnetic and the anisotropy is easy-plane ($A>0$). We use $J_1= 1.07$ meV, $J_{2}= 0.06$ meV, $J_{3}$= 0.24 meV, and $A$ = 0.0037 meV as obtained in Ref.\citenum{Rabindra2022PRM}. The interlayer exchange coupling is ferromagnetic\cite{Calder2021PRB} and we adapt an exponential decaying interlayer exchange of the form, $J_\perp(r)=J_\perp e^{-\alpha r}$ as considered for similar material systems\cite{Tong2018, Akram_NanoLett2024, Antao_arXiv2024}. For the remainder of the manuscript, we use $\alpha=5$ and investigate the phase diagram as a function of $J_\perp$. However our analysis indicates that our results are independent of $\alpha$, for $\alpha \gg 1$.  

The ground state of monolayer MnPSe$_3$ is a N\'eel AFM, characterized by an inplane order parameter ${\bf N}(\phi)=({\bf S}_A-{\bf S}_B)/2= \cos{\phi}\hat{x}+\sin{\phi}\hat{y}$ (in units of $|{\bf S}|=1$). Upon twisting, the interlayer exchange leads to frustration due to different local stacking patterns and AA and AB stacking regions form domains. Ref.~\citenum{Hejazi_PNAS2020} explores these multi-domain textures for twisted bilayer AFM systems in the continuum limit which is valid for weak interlayer exchange and large moir\'e period. In this limit, the ground state has two domains that are rotated by $90^o$ with respect to each other~\cite{Hejazi_PNAS2020}. In our analysis, we go beyond the continuum limit and we consider twisted bilayer lattice models. We obtain the ground state phase diagrams as a function of moir\'e period $L$ and interlayer coupling J$_\perp$ by solving the Landau-Lifshitz-Gilbert (LLG) equation\cite{1353448} for both layers (see Supporting Information for details). Since the AFM order parameter ${\bf N}(\phi)$ in a single domain only depends on $\phi$, we classify the ground states  by the maximum change in $\phi$, $(\Delta \phi)_{\text{max}}=(\phi_{AA}-\phi_{AB})$ between the AA and AB domains.  

The ground state is determined by the interplay between the energy cost due to domain wall and energy gain from the interlayer exchange. The interlayer energy is maximized when the order parameters of the two layers are parallel at AA stacking and antiparallel at AB stacking regions. Furthermore, the interlayer energy is proportional to area of the moir\'e unit cell $\sim L^2$, while the energy cost due to domain wall is proportional to the circumference of the domain which scales with $\sim L$\cite{Akram_NanoLett2021}. Therefore, the interlayer energy overcomes the domain wall cost for large $L$. Our phase diagram is presented in Fig.~\ref{Fig:1}(a) which shows that $(\Delta \phi)_{\text{max}}$ increases with increasing $L$ and J$_\perp$ until it reaches its maximum value 90$^\circ$, in agreement with the continuum calculations\cite{Hejazi_PNAS2020}. However, for smaller values of $L$ and J$_\perp$ the interplay of domain wall cost and interlayer energy lead to $(\Delta \phi)_{\text{max}}<90^\circ$. Fig.~\ref{Fig:1} (b) and (c) show the angle of order parameter in two layers. In Fig.~\ref{Fig:1}(b), $(\Delta \phi)_{\text{max}}<90^\circ$ and the order parameter changes continuously between AA and AB stacking regions for small $L$. However, for large $L$, $(\Delta \phi)_{\text{max}}\sim 90^\circ$ and the two domains are well separated by domain wall as shown in Fig.~\ref{Fig:1}(c).

\begin{figure}
\center
\includegraphics[width=0.7\textwidth]{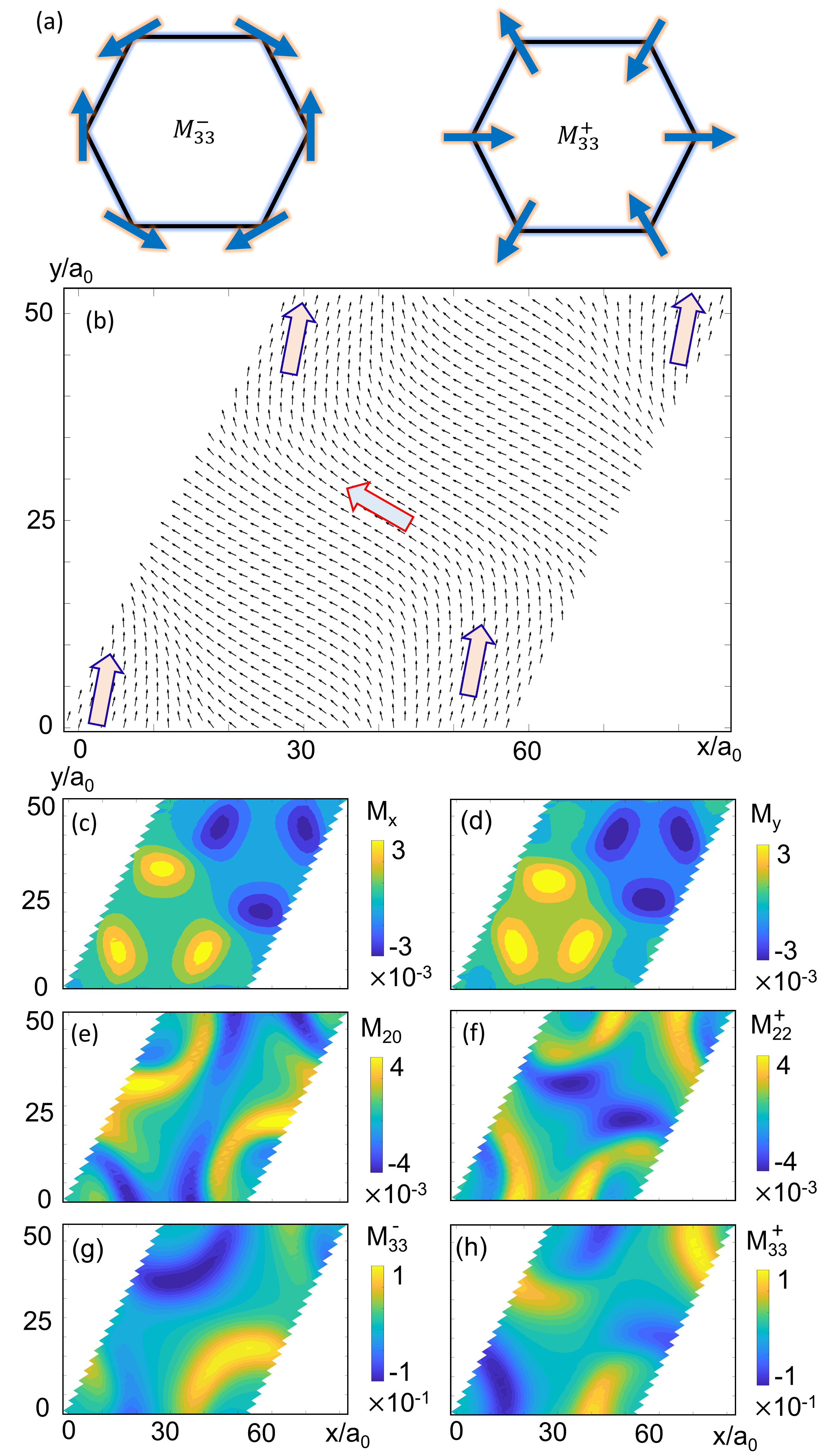} 
\caption{Multipole moments in a twisted bilayer of MnPSe$_3$. (a) Illustration of octupole moments $M_{33}^-$ and $M_{33}^+$. (b) Order parameter of layer-1 within a moir\'e unit cell at $L=60a_0$ and J$_{\perp}$=0.4 meV. (c,d) Components of magnetization of layer-1 within a moir\'e unit cell at $L=60 a_0$ and J$_{\perp}$=0.4 meV. (e,f) Quadrupole moments M$_{20}$ and M$_{22}^+$ of layer-1 within a moir\'e unit cell at $L=60a_0$ and J$_{\perp}$=0.4 meV. (g,h) Octupole moments M$_{33}^-$ and M$_{33}^+$ of layer-1 within a moir\'e unit cell at $L=60a_0$ and J$_{\perp}$=0.4 meV.}
\label{Fig:2_new}
\end{figure}

In a uniform N\'eel AFM, the multipolar order parameters including dipolar, quadrupolar and octupolar order vanish completely. However, multipolar orders are in general finite at domain walls since the order parameter varies. We define multipole moment per plaquette as $M_{lm}=\frac{1}{3}\sum_{i=1}^6 M_{lm,i}$ where the sum is over the hexagonal plaquette (see Supporting Information for details). The point group of MnPSe$_3$ is $D_{3d}$ and we consider the multipole moments for $D_{3d}$ given 
in Ref.~\citenum{PhysRevB.98.245129}.

\begin{figure}
\center
\includegraphics[width=1\textwidth]{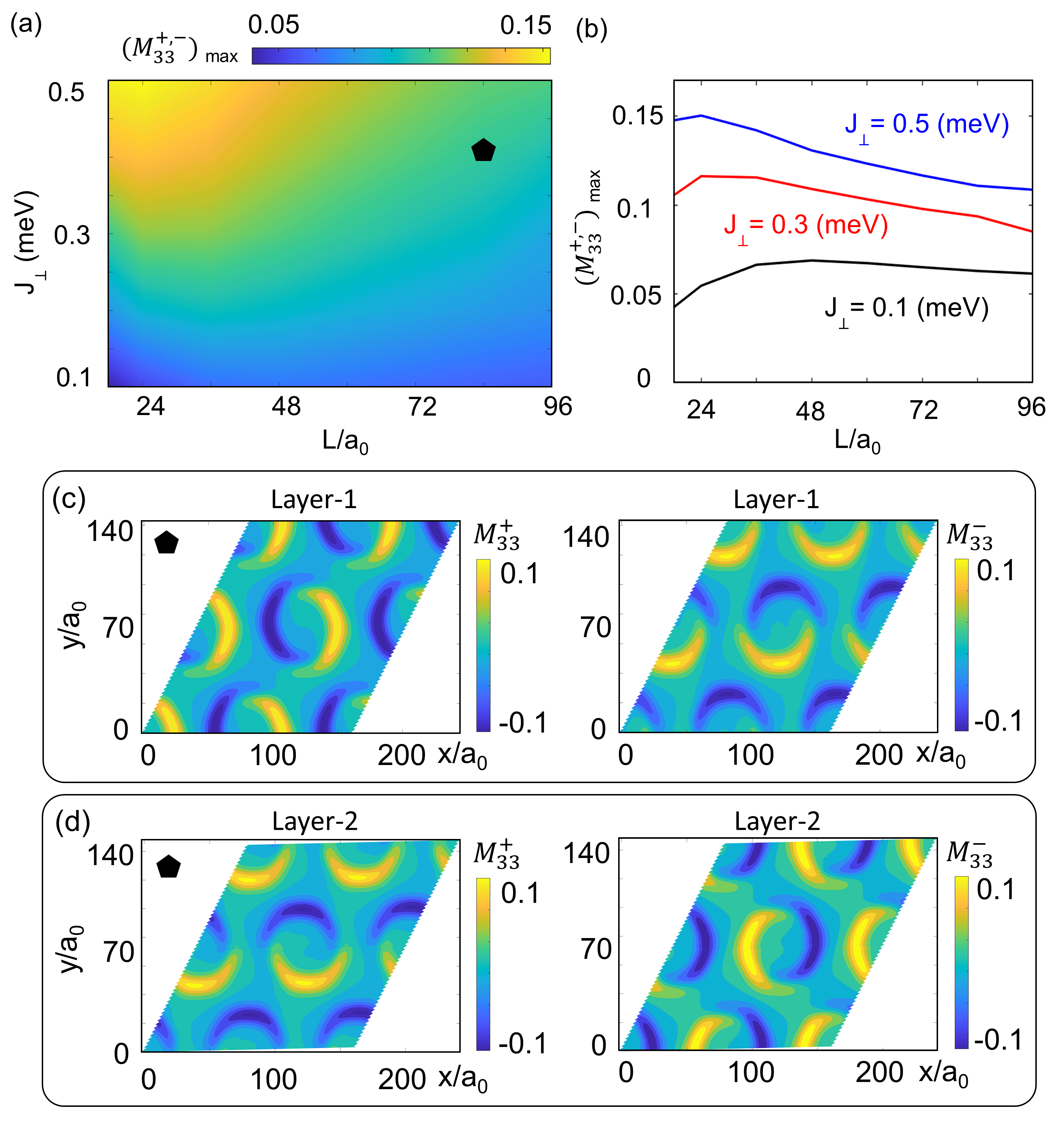} 
\caption{(a) Octupolar magnetic order in twisted bilayer MnPSe$_3$ as a function moir\'e period $L$ and interlayer coupling J$_{\perp}$. The color represents the amplitude of $M_{33}^+$ and $M_{33}^-$. (b) Amplitude of octupolar moments $(M_{33}^+)_{\text{max}}$ and $(M_{33}^-)_{\text{max}}$ as a function moir\'e period $L$ for J$_{\perp}=0.1, 0.2$ and 0.3 meV. (c) Octupolar moments $M_{33}^+$ and $M_{33}^-$ in layer-1 at $L=84a_0$ and J$_{\perp}$=0.4 meV for $2\times2$ moir\'e unit cells. (d) Octupolar moments $M_{33}^+$ and $M_{33}^-$ in layer-2 at $L=84a_0$ and J$_{\perp}$=0.4 meV for $2\times2$ moir\'e unit cells.}
\label{Fig:3}
\end{figure}

In Fig.~\ref{Fig:2_new}, we present the order parameter and multipole moments within a moir\'e unit cell. Our analysis indicates that the dipole and quadruple moments are negligible as shown in Fig.~\ref{Fig:2_new}(c,d,e,f). Similarly, octupole moments M$_{30}$, M$_{31}^-$, M$_{31}^+$ are also negligible while M$_{32}^-$ and M$_{32}^+$ are zero since $z$ and $S_z$ are zero for a 2D magnet with in-plane magnetization. However, octupole moments M$_{33}^+$ and M$_{33}^-$ are significant and they are of the order of $\sim 10^{-1}$ (in units of $\mu_B=1$, $|{\bf S}|=1$ and $a=1$) as shown in Fig.~\ref{Fig:2_new}(g,h). In Supporting Information, we show analytically that dipole moments are proportional to the third order in the variation of the order parameter, $|\boldsymbol{\nabla}\phi|^3$ at the domain wall. However, octupole moments $M_{33}^-\approx\bf{N}(\phi).\boldsymbol{\nabla}\phi$ and $M_{33}^+\approx\bf{N}(\phi+\pi/2).\boldsymbol{\nabla}\phi$ are proportional to first order in $|\boldsymbol{\nabla} \phi|$. In return, our analytical estimates show the underlying mechanism for the emergence of the octupolar order parameters, M$_{33}^{+/-}$ at the AFM domain walls and are in agreement with our numurical simulations.

\begin{figure}
\center
\includegraphics[width=1\textwidth]{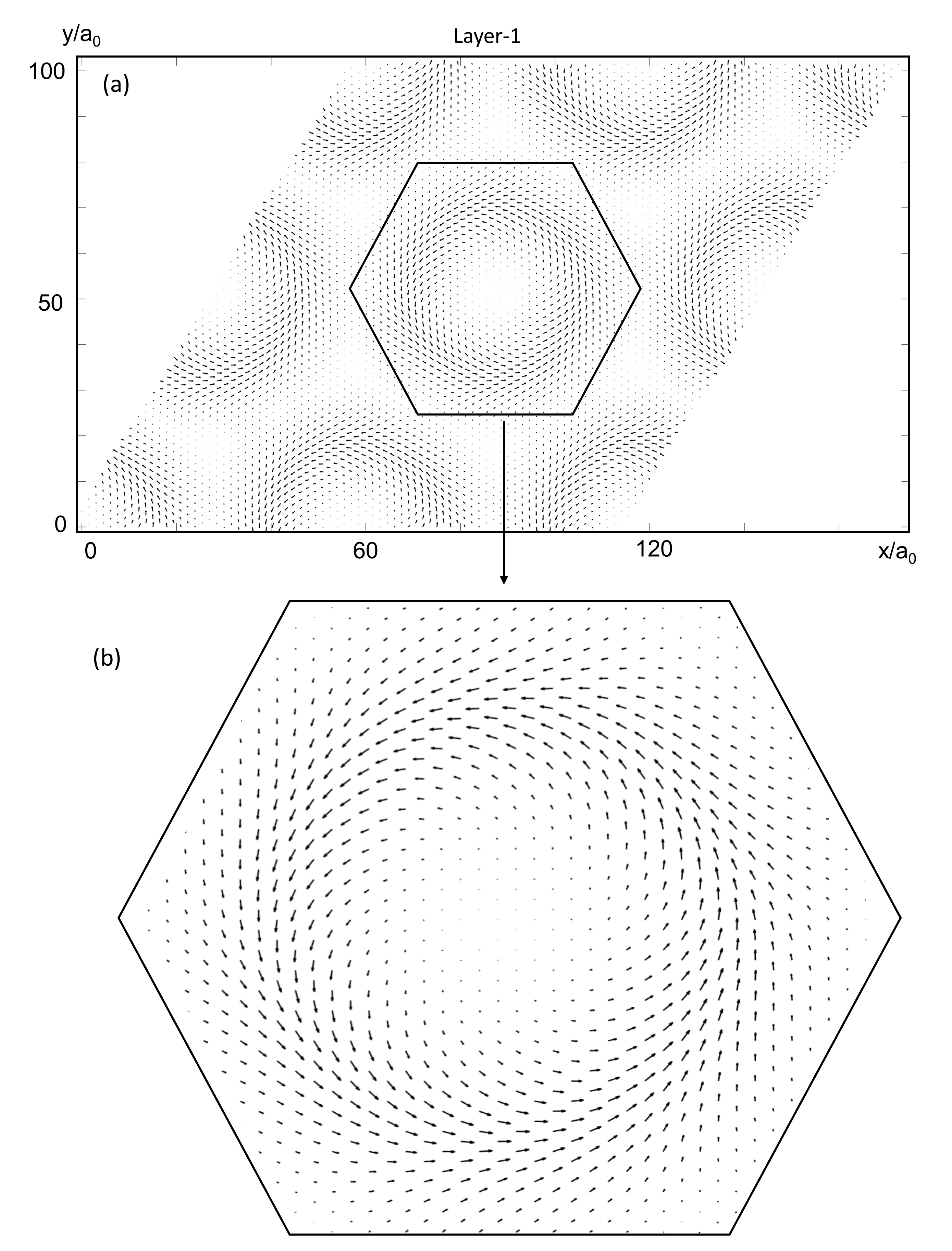} 
\caption{(a) Vortex crystal of octupole moments $M_{33}=[M_{33}^+,M_{33}^-]$ in layer-1 at $L=60a_0$ and J$_{\perp}$=0.4 meV for $2\times2$ moir\'e unit cells. (b) An expanded display of a octupolar vortex around a domain wall.}
\label{Fig:4}
\end{figure}

While the specific values of the octupolar order parameters depend on the interlayer exchange and moir\'e periodicity, the overall order of magnitude is about $10^{-1}$ for a wide parameter range as depicted in Fig.~\ref{Fig:3}(a). In addition, their maximum value increases with increasing J$_\perp$ since $(\Delta \phi)_{\text{max}}$ increases with J$_\perp$, consequently leading to an increase in $|\boldsymbol{\nabla}\phi|$. However, the amplitude initially increases with increasing $L$ until $(\Delta \phi)_{\text{max}}$ approaches $90^\circ$ after which it begins to decrease as $|\boldsymbol{\nabla}\phi|$ decreases for larger moir\'e periodicity, as depicted in Fig.~\ref{Fig:3}(b). Fig.~\ref{Fig:3} (c) and (d) display the octupole moments in $2\times 2$ moir\'e unit cells for both layers. For half of the domain wall, the moments are positive, while for the other half, they are negative. This disparity arises because the order parameter increases in one half of the domain wall and decreases in the other. The distribution of the two types of moments, $M_{33}^+$ and $M_{33}^-$, are related by a $90^\circ$ rotation within each layer, as shown in Fig. 3(c,d). Moreover, $M_{33}^+$ and $M_{33}^-$ in layer-1 are related to $M_{33}^+$ and $M_{33}^-$ in layer-2 by a $90^\circ$ rotation respectively. Consequently, there is no net octupole moment at either layer.

\begin{figure}[t]
\center
\includegraphics[width=0.8\textwidth]{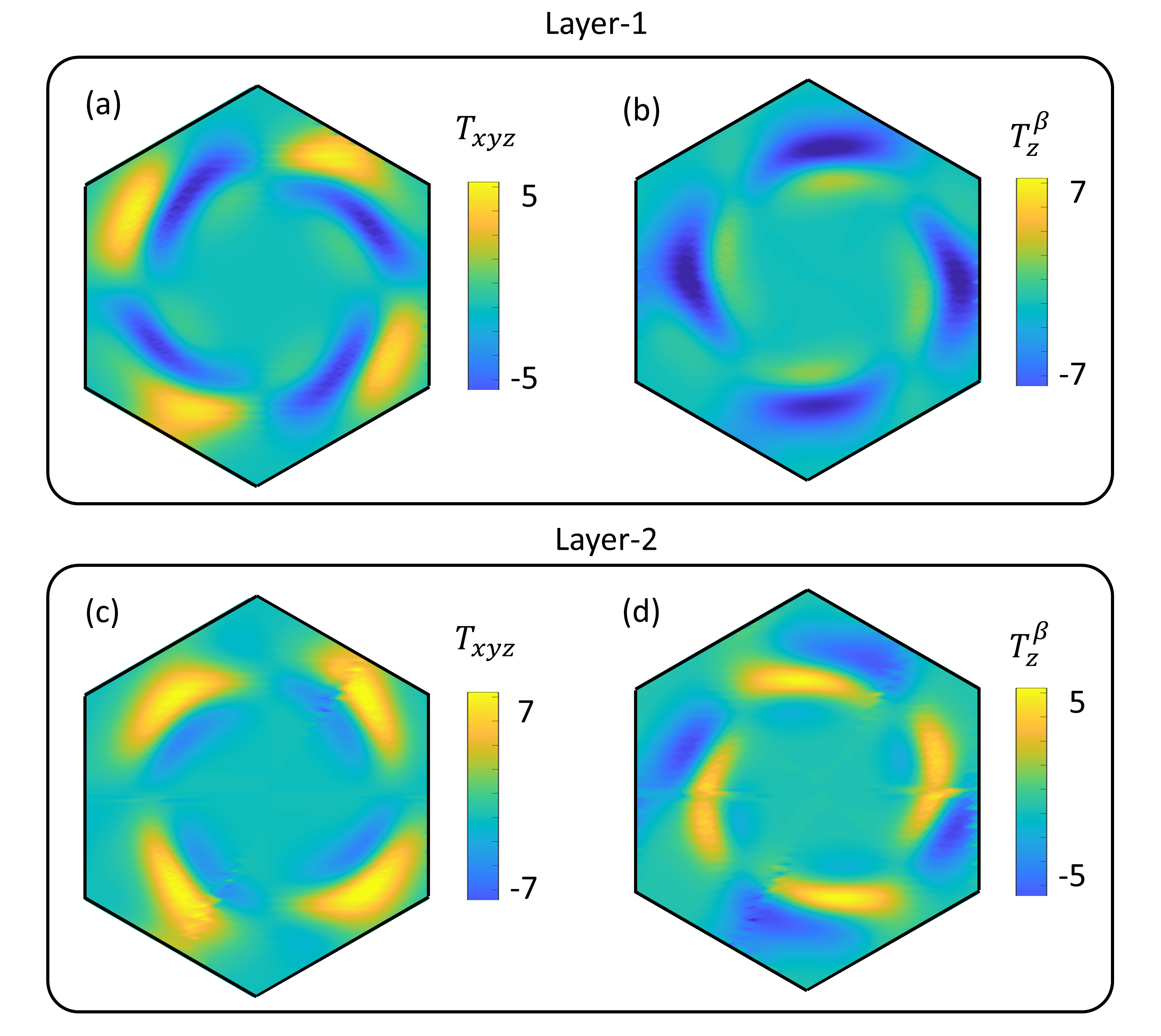} 
\caption{(a,b) Toroidal octupole moments $T_{xyz}$ and $T_{z}^{\beta}$ in the moir\'e unit cell of layer-1 at $L=84a_0$ and J$_{\perp}$=0.4 meV. (c,d) Toroidal octupole moments $T_{xyz}$ and $T_{z}^{\beta}$ in the moir\'e supercell of layer-2 at $L=84a_0$ and J$_{\perp}$=0.4 meV.}
\label{Fig:5}
\end{figure}

In order to gain more intuition about this magnetic state, we represent the octupole moments $M_{33}^-$ and $M_{33}^+$ by a vector $\boldsymbol{M}_{33}=[M_{33}^-,M_{33}^+]$ where $M_{33}^-$ and $M_{33}^+$ are x and y-components respectively. Fig.~\ref{Fig:4}(a) illustrates the vector $\boldsymbol{M}_{33}$ in $2\times2$ moir\'e unit cells. $\boldsymbol{M}_{33}$ forms vortices around the domain walls and leads to one vortex per moir\'e unit cell. The formation of the vortices can be deduced from Fig.~\ref{Fig:3}(c) and (d) and we present analytical derivation in the Supporting Information. 
%
This winding of the magnetic octupole moments gives rise to two separate magnetotoroidal octupole moments denoted as $T_{xyz}$ and $T_z^\beta$, as shown in Fig.~\ref{Fig:5}.
Unlike octupole moments, there is a net toroidal octupole moments within each moir\'e supercell. Additionally, the net toroidal octupole moments depend on the orientation of the order parameter in the AA and AB regions, and the values of these moments differ between the two layers: $T_{xyz}=7.01\times 10^2$ and $T_{z}^\beta=-2.96\times10^3$ on layer-1, and $T_{xyz}=3.03\times10^3$ and $T_{z}^\beta=-191$ on layer-2.

Macroscopic multipoles induced by ferro-ordering of microscopic unit cell multipoles in materials can be probed by exploiting the macroscopic response tensors to gain insight into the symmetry of materials \cite{PhysRevB.98.245129}. For example, the electric and magnetic dipoles give rise to optical rotations through natural optical activity and spontaneous Faraday effect respectively, both of which correspond to specific components of the optical permittivity tensor.\cite{Newnham2005} 
However, higher rank multipoles are usually harder to observe. In situations where the irrep of the point group that corresponds to a higher rank multipole is identical to that of a lower rank multipole, this can be exploited to observe the higher rank multipole's effects. For example, in the antiferromagnetic Mn$_3$Ir and Mn$_3$Z (where Z=Sn, Ge) the magnetic dipole and octupoles have components that transform as the same irrep. While the magnetic order in these compounds are primarily octupolar in nature, this group theoretical condition implies that there is also a net magnetic dipole moment. (Hence, these materials can be classified as weak ferromagnetic altermagnets.) The anomalous Hall effects in these compounds are proportional to the octupolar order parameter since there is no symmetry condition that distinguished the octupole from a dipole\cite{PhysRevB.98.245129}. Experimental observation of a very sizable anomalous Hall effect has been reported in these compounds Mn$_3$Ir and Mn$_3$Z (Z=Sn, Ge)\cite{PhysRevLett.112.017205,PhysRevApplied.5.064009,doi:10.1126/sciadv.1501870,Nakatsuji_large_anomalous_Hall_Nature}. 

In twisted bilayer MnPSe$_3$, the octupole moments neither belong to the same irrep as dipoles, nor do they sum up to a nonzero value in the moir\'e unit cell since they have opposite signs on either side of the domain walls. There is no reason to expect that an anomalous Hall effect emerge in these systems, and one needs to consider higher rank tensors to detect the order present in this system. 
Point group representations allow a systematic means to predict what tensors can be used to measure a specific order parameter. For example, the rank-4 elastoresistivity (or piezoconductivity) tensor can be used to directly probe any inversion and time-reversal even order, such as the ferrorotational (ferro-electro-toroidal) order \cite{Dayroberts2024}. 
In a similar vein, the vortices formed by the octupole moments and the resultant toroidal octupole moment can be probed by macroscopic means using a suitable tensorial response. The doublet formed by $T_{xyz}$ and $T_z^\beta$ transform as the $E_{2u}^-$ irrep of the point group $6/mmm$. This irrep is odd under both time-reversal and inversion operations, which determines the limits of experimental probes that can be used. Typically, this type of orders odd under both time and spatial inversion give rise to magneto-electric effects, as well as magneto-optical effects. 

The linear magneto-electric effect tensor $\alpha_{ij}$ is defined as the linear dependence of the components of magnetization $M_i$ on the applied electric field components $E_j$ through $M_i=\alpha_{ij} E_j$.\cite{Birol2012} Using the so-called Jahn symbols, $\alpha$ can be represented as $aeV^2$, which means that it is a rank-2 tensor that is not either symmetric or antisymmetric, and is inversion odd ($e$), as well as time reversal odd ($a$).\cite{Gallego_AC2019} Its 9 independent components form a 9-dimensional  reducible representation of $6/mmm$, which is \begin{equation}
\Gamma^\alpha=\Gamma^{aeV^2}= 2A_{1u}^-+A_{2u}^-+2E_{1u}^-+E_{2u}^-. 
\end{equation}
The two components of $\alpha$ that transform as the two basis functions of the $E_{2u}^-$ irrep are $\left[ \left(\alpha_{xx}-\alpha_{yy}\right), \left(\alpha_{xy}+\alpha_{yx}\right)\right]$. In other words, the octupolar multipole gives rise to a traceless in-plane magnetoelectric effect tensor, which corresponds to an in-plane electric (magnetic) field inducing an in-plane magnetization (electric polarization). This effect has both longitudinal and  transverse components. 

One can also consider rank-3 tensors whose expansion contain the $E_{2u}^-$ irrep of point group $6/mmm$. A rank-3 polar tensor is odd under inversion, and if it is symmetric under the exchange of first two indices, it is represented by the Jahn symbol $a[V^2]V$. (If the tensor is symmetric under the exchange of the last two indiced, the Jahn symbol would be $aV[V^2]$). The letter $a$ in the Jahn symbol means that the tensor is odd under time reversal operation.\cite{Gallego_AC2019} Both of these Jahn symbols lead to the expansion 
\begin{equation}
    \Gamma^{a[V^2]V}= A_{1u}^-+3 A_{2u}^-+B_{1u}^-+B_{2u}^-+4E_{1u}^-+2E_{2u}^- . 
\end{equation}
$E_{2u}^-$ appears twice in this expansion, which means that there are two pairs of elements of the tensor that would emerge and be linearly proportional to the magneto-toroidal-octupolar moment in the material. (These components are listed in the Supporting Information.) 

To the best of our knowledge, there are no static response tensors that are easy to experimentally measure, except the `piezotoroidic effect tensor' which connects stress (or strain) on a material with a magneto-toroidal dipole. Considering the optical effects provides more options. The spontaneous gyrotropic birefringence, which is similar to the spontaneous Faraday effect except that it changes sign when the propagation direction of light is reversed, is an example of optical effects that transform as $aV[V^2]$. While this effect is predicted to exist in certain AFMs as well as in cuprate superconductors\cite{Varma_EPL2014}, it is challenging to observe experimentally in commonly studied materials systems.\cite{Iguchi2021} We propose twisted antiferromagnets that give rise to a macroscopic magnetotoroidal octupole as a novel platform to search for this effect. Certain higher order optical susceptibilities, discussed in Ref.~\citenum{Gallego_AC2019}, can also be used for this aim, since they also have the same Jahn symbol. 

In passing, we note that while many different higher order magnetic multipoles have been associated with the emerging class of so-called altermagnetic materials \cite{Bhowal2024, Fernandes2024}, magnetotoroidal octupoles do not lead to altermagnetism because they are odd under spatial inversion.

In conclusion, we have shown that the interplay of stacking-dependent interlayer exchange and twist angle play an important role in determining the magnetic phases of MnPSe$_3$. Specifically, we have demonstrated the existence of a two-domain phase, where the difference in angle of order parameters between the two domains increases with increasing interlayer coupling and moir\'e period until it reaches 90$^\circ$. Additionally, our findings show that octupole moments M$_{33}^+$ and M$_{33}^-$ are significant at domain walls, while dipole and quadrupole moments are negligible. These octupole moments form vortex-like structure at the scale of the Moire cell and give rise to net magnetotoroidal octupole moments, $T_{xyz}$ and $T_z^\beta$. We propose linear diagonal magnetoelectric effect and spontaneous gyrotropic birefringence as possible means to probe these emergent moments.

\section{Associated Content}

\noindent {\bf Supporting Information}

\noindent {Details of the atomistic simulations, Multipole moments, magnetization and octupole moments at an antiferromagnetic domain wall, octupolar vortex formation, representation of $a[V^2]V$ in point group $6/mmm$ ($D_{6h}$) (pdf)}.

\section{Author Information}
\noindent {\bf Corresponding Author:}
Onur Erten -- Email: onur.erten@asu.edu

\noindent {\bf Notes:} The authors declare no competing financial interest.

\section{Acknowledgements}
We thank Cristian Batista and Liuyan Zhao for fruitful discussions.
The work at ASU was supported by the U.S. Department of Energy, Office of Science, Office of Basic Energy Sciences, Material Sciences and Engineering Division under Award Number DE-SC002524. FY and TB were supported by the NSF CAREER grant DMR-2046020. We thank ASU Research Computing Center for high-performance computing resources.

\bibliography{references}
\end{document}


\title{Supporting Information for ``Octupolar vortex crystal and toroidal moment in twisted bilayer MnPSe$_3$"}

\author{Muhammad Akram}
\affiliation{Department of Physics, Arizona State University, Tempe, AZ 85287, USA}
\affiliation{Department of Physics, Balochistan University of Information Technology, Engineering and Management Sciences (BUITEMS), Quetta 87300, Pakistan}
\author{Fan Yang}
\affiliation{Department of Chemical Engineering and Materials Science, University of Minnesota, Minneapolis, MN 55455, USA}
\author{Turan Birol}
\affiliation{Department of Chemical Engineering and Materials Science, University of Minnesota, Minneapolis, MN 55455, USA}
\author{Onur Erten}
\affiliation{Department of Physics, Arizona State University, Tempe, AZ 85287, USA}


\maketitle

\section{Details of the atomistic simulations} \label{app:LLG}
To determine the ground state of the Hamiltonian, we solve the Landau-Lifshitz-Gilbert (LLG) equations: 
\begin{eqnarray} \label{eq:E1}
\frac{d\textbf{S}}{dt}=-\gamma \textbf{S}\times \textbf{B}^{\rm eff}+\alpha \textbf{S} \times \frac{d\textbf{S}}{dt} \, ,
\end{eqnarray}
where $\textbf{B}^{\rm eff}=-\delta H/\delta \textbf{S}$ represents the effective magnetic field, $\gamma$ is the gyromagnetic ratio and $\alpha$ is Gilbert damping coefficient. We solve the LLG equations self-consistently, imposing the constraint $|{\bf{S}}|=1$ and applying periodic boundary conditions. The semi-implicit midpoint method 
is implemented in MATLAB software. We explored several random initial states for a specific interlayer exchange coupling and twist angle, and picked the lowest energy configuration after achieving convergence.

\section{Multipole moments}
The point group of MnPSe$_3$ is $D_{3d}$ and multipole moments for this group in real space basis are:

\begin{eqnarray}
\text{Dipole moment: }\textbf{M}&=&\textbf{S} \\
\text{Quadrupole moments: } M_{20}&=&2zS_z-xS_x-yS_y \\
M_{21}^-&=&yS_z+zS_y \\
M_{21}^+&=&zS_x+xS_z \\
M_{22}^-&=&xS_y+yS_x \\
M_{22}^+&=&xS_x-yS_y \\
\text{Octupole moments: } M_{30}&=&S_z \\
\{M_{31}^+,M_{31}^-\}&=&\{S_x,S_y\} \\
\{M_{32}^+,M_{32}^-\}&=&\{(x^2-y^2)S^z,xyS^z\}, \\
&&\{z(xS_x-yS_y),z(yS_x+xS_y)\}\\
M_{33}^+&=&(x^2-y^2)S^x-2xyS^y \\
M_{33}^-&=&(x^2-y^2)S^y+2xyS^x\\
\text{Toroidal moments: } \textbf{t}_3 &=& \textbf{r}\times\textbf{S}\\
\text{Toroidal octupole moments: } \\
T_{xyz}&=&\sqrt{15}[yzt_3^x+zxt_3^y+xyt_3^z] \\
T_{z}^\beta&=&\sqrt{15}[\frac{1}{2}(x^2-y^2)t_3^z+z(zt_3^x-yt_3^y))]
\end{eqnarray}

\begin{figure}
\center
\includegraphics[width=0.9\textwidth]{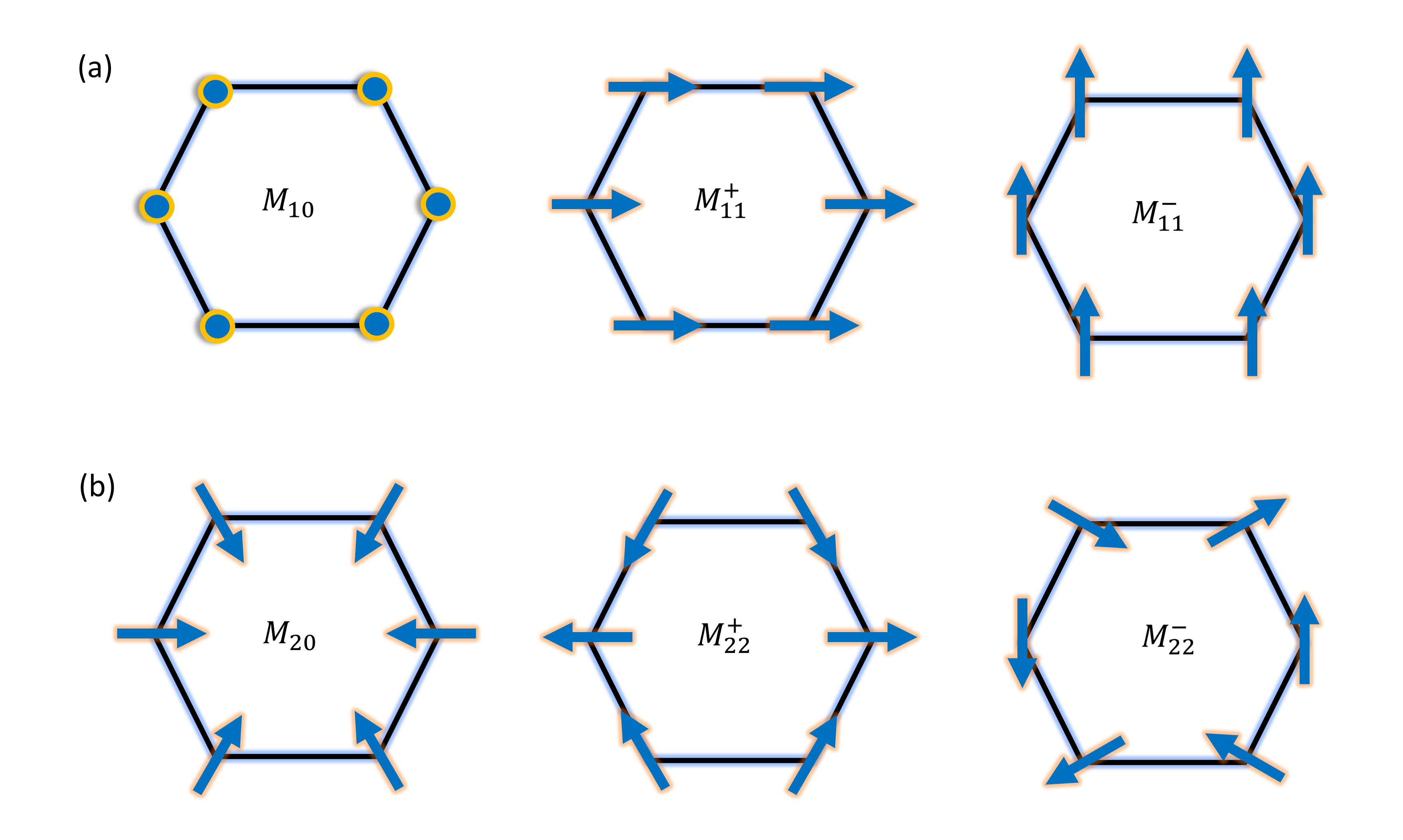} 
\caption{Real space representation of dipole and quadrupole moments. (a) Real space representation of dipole moments, where spheres indicate the z-component of magnetization and arrows illustrate the in-plane component of magnetization. (b) Real space representation of quadrupole moments.}
\label{Fig:dip_quad}
\end{figure}

\section{\label{sec:details} Magnetization and octupole moments at an antiferromagnetic domain wall}

\begin{figure}
\center
\includegraphics[width=0.7\textwidth]{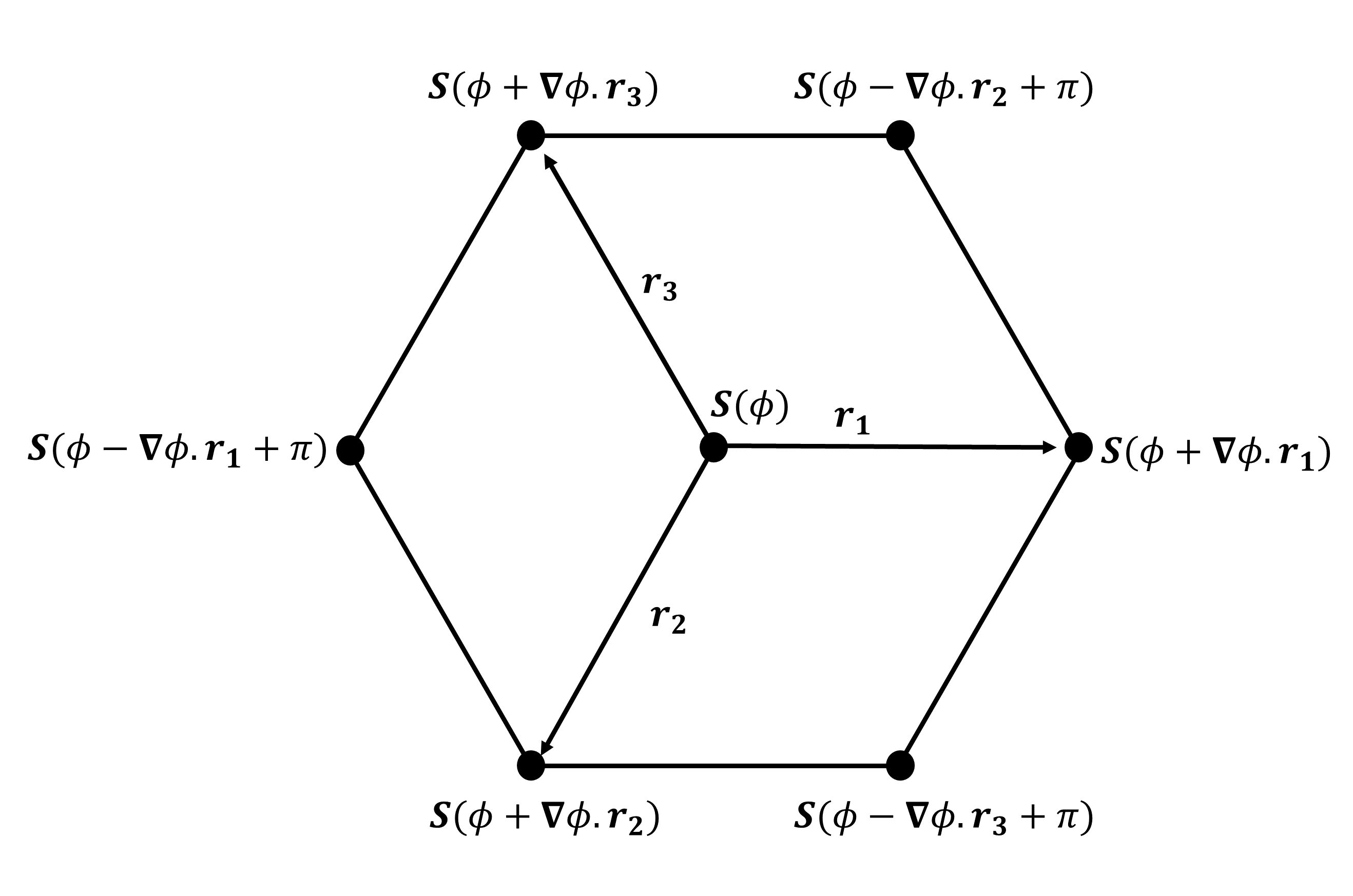} 
\caption{Magnetization at the corners of a hexagon in a domain wall.}
\label{Fig:mag_domain_wall}
\end{figure}

The magnetization per hexagon M is defined as the sum of the spins at the vertices of each hexagon, divided by three, since each vertex contributes to three hexagons. To calculate the magnetization in the domain wall, we examine a hexagon, as shown in Fig.\ref{Fig:mag_domain_wall}, where the order parameter changes in space.

\begin{eqnarray}
    \textbf{M}&=&\frac{1}{3}( \textbf{S}_1+\textbf{S}_2+\textbf{S}_3+\textbf{S}_4+\textbf{S}_5+\textbf{S}_6)\\    
    &=&\frac{1}{3}(\textbf{S}(\phi+\bm{\nabla}\phi.\textbf{r}_1)+\textbf{S}(\phi+\bm{\nabla}\phi.\textbf{r}_2)+\textbf{S}(\phi+\bm{\nabla}\phi.\textbf{r}_3)\notag\\    
    && +\textbf{S}(\phi-\bm{\nabla}\phi.\textbf{r}_1+\pi)+\textbf{S}(\phi-\bm{\nabla}\phi.\textbf{r}_2+\pi)+\textbf{S}(\phi-\bm{\nabla}\phi.\textbf{r}_3+\pi)) \notag\\    
    &=&\frac{1}{3}(\textbf{S}(\phi+\bm{\nabla}\phi.\textbf{r}_1)+\textbf{S}(\phi+\bm{\nabla}\phi.\textbf{r}_2)+\textbf{S}(\phi+\bm{\nabla}\phi.\textbf{r}_3)\notag\\    
    && -\textbf{S}(\phi-\bm{\nabla}\phi.\textbf{r}_1)-\textbf{S}(\phi-\bm{\nabla}\phi.\textbf{r}_2)-\textbf{S}(\phi-\bm{\nabla}\phi.\textbf{r}_3)) \notag\\    
    \label{eq:Mag}
\end{eqnarray}

Consider $\textbf{S}(\phi)=[S^x,S^y]=[\cos{\phi}, \sin{\phi}]$
\begin{eqnarray}
    M^x&=&\frac{1}{3}(S_1^x+S_2^x+S_3^x+S_4^x+S_5^x+S_6^x)\notag\\
    &=&\frac{1}{3}(\cos{(\phi+\bm{\nabla}\phi.\textbf{r}_1)}+\cos{(\phi+\bm{\nabla}\phi.\textbf{r}_2)}+\cos{(\phi+\bm{\nabla}\phi.\textbf{r}_3)}\notag\\
    && -\cos{(\phi-\bm{\nabla}\phi.\textbf{r}_1)}-\cos{(\phi-\bm{\nabla}\phi.\textbf{r}_2)}-\cos{(\phi-\bm{\nabla}\phi.\textbf{r}_3)})\notag\\
    &=&\frac{1}{3}(\,-2\sin{\phi}\sin{(\bm{\nabla}\phi.\textbf{r}_1)}-2\sin{\phi}\sin{(\bm{\nabla}\phi.\textbf{r}_2)}-2\sin{\phi}\sin{(\bm{\nabla}\phi.\textbf{r}_3)})\notag\\
    &=&\frac{-2\sin{\phi}}{3}(\,\sin{(\bm{\nabla}\phi.\textbf{r}_1)}+\sin{(\bm{\nabla}\phi.\textbf{r}_2)}+\sin{(\bm{\nabla}\phi.\textbf{r}_3)})\notag\\ 
    &=&\frac{-2\sin{\phi}}{3}(\,(\bm{\nabla}\phi.\textbf{r}_1)-\frac{(\bm{\nabla}\phi.\textbf{r}_1)^3}{3!}+... +(\bm{\nabla}\phi.\textbf{r}_2)-\frac{(\bm{\nabla}\phi.\textbf{r}_2)^3}{3!}+... +(\bm{\nabla}\phi.\textbf{r}_3)-\frac{(\bm{\nabla}\phi.\textbf{r}_3)^3}{3!}+...)\notag\\ 
    &=&\frac{-2\sin{\phi}}{3}(\,(\textbf{r}_1+\textbf{r}_2+\textbf{r}_3).\bm{\nabla}\phi+\mathcal{O}(3))\notag\\ 
    &=&\frac{-2\sin{\phi}}{3}(\,\mathcal{O}(3)) 
\end{eqnarray}

\begin{eqnarray}
    M^y&=&\frac{1}{3}(S_1^y+S_2^y+S_3^y+S_4^y+S_5^y+S_6^y)\notag\\
    &=&\frac{1}{3}(\sin{(\phi+\bm{\nabla}\phi.\textbf{r}_1)}+\sin{(\phi+\bm{\nabla}\phi.\textbf{r}_2)}+\sin{(\phi+\bm{\nabla}\phi.\textbf{r}_3)}\notag\\
    && -\sin{(\phi-\bm{\nabla}\phi.\textbf{r}_1)}-\sin{(\phi-\bm{\nabla}\phi.\textbf{r}_2)}-\sin{(\phi-\bm{\nabla}\phi.\textbf{r}_3)})\notag\\
    &=&\frac{1}{3}(\,2\cos{\phi}\sin{(\bm{\nabla}\phi.\textbf{r}_1)}+2\cos{\phi}\sin{(\bm{\nabla}\phi.\textbf{r}_2)}+2\cos{\phi}\sin{(\bm{\nabla}\phi.\textbf{r}_3)})\notag\\
    &=&\frac{2\cos{\phi}}{3}(\,\sin{(\bm{\nabla}\phi.\textbf{r}_1)}+\sin{(\bm{\nabla}\phi.\textbf{r}_2)}+\sin{(\bm{\nabla}\phi.\textbf{r}_3)})\notag\\ 
    &=&\frac{2\cos{\phi}}{3}(\,(\bm{\nabla}\phi.\textbf{r}_1)-\frac{(\bm{\nabla}\phi.\textbf{r}_1)^3}{3!}+... +(\bm{\nabla}\phi.\textbf{r}_2)-\frac{(\bm{\nabla}\phi.\textbf{r}_2)^3}{3!}+... +(\bm{\nabla}\phi.\textbf{r}_3)-\frac{(\bm{\nabla}\phi.\textbf{r}_3)^3}{3!}+...)\notag\\ 
    &=&\frac{2\cos{\phi}}{3}(\,(\textbf{r}_1+\textbf{r}_2+\textbf{r}_3).\bm{\nabla}\phi+\mathcal{O}(3))\notag\\ 
    &=&\frac{2\cos{\phi}}{3}(\,\mathcal{O}(3)) 
\end{eqnarray}

The octopole moments for ${D_{3d}}$ point group in real space basis are $M_{33}^+=(x^2-y^2)S^x-2xyS^y$ and $M_{33}^-=(x^2-y^2)S^y+2xyS^x$.
Here $(x,y)$ are the position coordinates of spin with respect to center of hexagon and $(S_x,S_y)$ are components of spin. The octopole moments per hexagon are defined as
\begin{eqnarray}
    M_{33}^+&=&\frac{1}{3}(M_{33,1}^++M_{33,2}^++M_{33,3}^++M_{33,4}^++M_{33,5}^++M_{33,6}^+)\notag\\
    &=&\frac{1}{3}((x_1^2-y_1^2)S_1^x-2x_1y_1S_1^y+(x_2^2-y_2^2)S_2^x-2x_2y_2S_2^y+(x_3^2-y_3^2)S_3^x-3x_3y_3S_3^y\notag\\
    &&+(x_4^2-y_4^2)S_4^x-2x_4y_4S_4^y+(x_5^2-y_5^2)S_5^x-2x_5y_5S_5^y+(x_6^2-y_6^2)S_6^x-2x_6y_6S_6^y)\notag\\
    &=&\frac{1}{3}((x_1^2-y_1^2)S_1^x+(x_2^2-y_2^2)S_2^x+(x_3^2-y_3^2)S_3^x+(x_4^2-y_4^2)S_4^x+(x_5^2-y_5^2)S_5^x+(x_6^2-y_6^2)S_6^x\notag\\
    &&-2x_1y_1S_1^y-2x_2y_2S_2^y-2x_3y_3S_3^y-2x_4y_4S_4^y-2x_5y_5S_5^y-2x_6y_6S_6^y)
    \label{eq:M_33_plus}
\end{eqnarray}

Consider
\begin{eqnarray}
    &=&(x_1^2-y_1^2)S_1^x+(x_2^2-y_2^2)S_2^x+(x_3^2-y_3^2)S_3^x+(x_4^2-y_4^2)S_4^x+(x_5^2-y_5^2)S_5^x+(x_6^2-y_6^2)S_6^x\notag\\
    &=&S_1^x+S_2^x-\frac{1}{2}S_3^x-\frac{1}{2}S_4^x-\frac{1}{2}S_5^x-\frac{1}{2}S_6^x\notag\\
    &=&\cos{(\phi+\bm{\nabla}\phi.\textbf{r}_1)}-\cos{(\phi-\bm{\nabla}\phi.\textbf{r}_1)}\notag\\
    &&-\frac{1}{2}(\cos{(\phi+\bm{\nabla}\phi.\textbf{r}_2)}-\cos{(\phi-\bm{\nabla}\phi.\textbf{r}_2)}+\cos{(\phi+\bm{\nabla}\phi.\textbf{r}_3)}
     -\cos{(\phi-\bm{\nabla}\phi.\textbf{r}_3)})\notag\\
     &=& -2\sin{\phi}\sin{(\bm{\nabla}\phi.\textbf{r}_1)}-\frac{1}{2}(-2\sin{\phi}\sin{(\bm{\nabla}\phi.\textbf{r}_2)}-2\sin{\phi}\sin{(\bm{\nabla}\phi.\textbf{r}_3)})\notag\\
     &=& \sin{\phi}(-2\sin{(\bm{\nabla}\phi.\textbf{r}_1)}+\sin{(\bm{\nabla}\phi.\textbf{r}_2)}+\sin{(\bm{\nabla}\phi.\textbf{r}_3)})\notag\\
     &=& \sin{\phi}(-2\bm{\nabla}\phi.\textbf{r}_1+\bm{\nabla}\phi.\textbf{r}_2+\bm{\nabla}\phi.\textbf{r}_3+\mathcal{O}(3))\notag\\
     &\approx& \sin{\phi}(-2\textbf{r}_1+\textbf{r}_2+\textbf{r}_3).\bm{\nabla}\phi\notag\\
     &\approx& \sin{\phi}(-2\frac{\partial \phi}{\partial x}-\frac{1}{2}\frac{\partial \phi}{\partial x}-\frac{\sqrt{3}}{2}\frac{\partial \phi}{\partial y}-\frac{1}{2}\frac{\partial \phi}{\partial x}+\frac{\sqrt{3}}{2}\frac{\partial \phi}{\partial y})\notag\\
     &\approx& -3\sin{\phi}\frac{\partial \phi}{\partial x} 
     \label{eq:M_33_plus1}
\end{eqnarray}

Now consider
\begin{eqnarray}
    &=&-2x_1y_1S_1^y-2x_2y_2S_2^y-2x_3y_3S_3^y-2x_4y_4S_4^y-2x_5y_5S_5^y-2x_6y_6S_6^y\notag\\
    &=&0.S_1^y+0.S_2^y-\frac{\sqrt{3}}{2}(S_3^y+S_4^y)+\frac{\sqrt{3}}{2}(S_5^y+S_6^y)\notag\\
    &=&-\frac{\sqrt{3}}{2}(\sin{(\phi+\bm{\nabla}\phi.\textbf{r}_2)}-\sin{(\phi-\bm{\nabla}\phi.\textbf{r}_2)}+\frac{\sqrt{3}}{2}(\sin{(\phi+\bm{\nabla}\phi.\textbf{r}_3)}
     -\sin{(\phi-\bm{\nabla}\phi.\textbf{r}_3)})\notag\\
     &=& -\frac{\sqrt{3}}{2}(2\cos{\phi}\sin{(\bm{\nabla}\phi.\textbf{r}_2)})+\frac{\sqrt{3}}{2}(2\sin{\phi}\sin{(\bm{\nabla}\phi.\textbf{r}_3)})\notag\\
     &=& \sqrt{3}\cos{\phi}(-\sin{(\bm{\nabla}\phi.\textbf{r}_2)}+\sin{(\bm{\nabla}\phi.\textbf{r}_3)})\notag\\
     &=& \sqrt{3}\cos{\phi}(-\bm{\nabla}\phi.\textbf{r}_2+\bm{\nabla}\phi.\textbf{r}_3+\mathcal{O}(3))\notag\\
     &\approx& \sqrt{3}\cos{\phi}(-\textbf{r}_2+\textbf{r}_3).\bm{\nabla}\phi\\
     &\approx& \sqrt{3}\cos{\phi}(\frac{1}{2}\frac{\partial \phi}{\partial x}+\frac{\sqrt{3}}{2}\frac{\partial \phi}{\partial y}-\frac{1}{2}\frac{\partial \phi}{\partial x}+\frac{\sqrt{3}}{2}\frac{\partial \phi}{\partial y})\notag\\
     &\approx& 3\cos{\phi}\frac{\partial \phi}{\partial y}
     \label{eq:M_33_plus2}
\end{eqnarray}
Using \ref{eq:M_33_plus1} and \ref{eq:M_33_plus2} in \ref{eq:M_33_plus} we obtain
\begin{eqnarray}
M_{33}^+&\approx&-\sin{\phi}\frac{\partial \phi}{\partial x}+\cos{\phi}\frac{\partial \phi}{\partial y}\notag\\
&\approx&\cos{(\phi+\pi/2)}\frac{\partial \phi}{\partial x}+\sin{(\phi+\pi/2)}\frac{\partial \phi}{\partial y}\notag\\
&\approx & \bf{N}(\phi+\pi/2).\bf{\nabla}\phi
\end{eqnarray}

Now
\begin{eqnarray}
    M_{33}^-&=&\frac{1}{3}(M_{33,1}^-+M_{33,2}^-+M_{33,3}^-+M_{33,4}^-+M_{33,5}^-+M_{33,6}^-)\notag\\
    &=&\frac{1}{3}((x_1^2-y_1^2)S_1^y+2x_1y_1S_1^x+(x_2^2-y_2^2)S_2^y+2x_2y_2S_2^x+(x_3^2-y_3^2)S_3^y+3x_3y_3S_3^x\notag\\
    &&+(x_4^2-y_4^2)S_4^y+2x_4y_4S_4^x+(x_5^2-y_5^2)S_5^y+2x_5y_5S_5^x+(x_6^2-y_6^2)S_6^y+2x_6y_6S_6^x)\notag\\
    &=&\frac{1}{3}((x_1^2-y_1^2)S_1^y+(x_2^2-y_2^2)S_2^y+(x_3^2-y_3^2)S_3^y+(x_4^2-y_4^2)S_4^y+(x_5^2-y_5^2)S_5^y+(x_6^2-y_6^2)S_6^y\notag\\    &&2x_1y_1S_1^x+2x_2y_2S_2^x+2x_3y_3S_3^x+2x_4y_4S_4^x+2x_5y_5S_5^x+2x_6y_6S_6^x)
    \label{eq:M_33_minus}
\end{eqnarray}

Consider

\begin{eqnarray}
    &=&(x_1^2-y_1^2)S_1^y+(x_2^2-y_2^2)S_2^y+(x_3^2-y_3^2)S_3^y+(x_4^2-y_4^2)S_4^y+(x_5^2-y_5^2)S_5^y+(x_6^2-y_6^2)S_6^y\notag\\
    &=&S_1^y+S_2^y-\frac{1}{2}S_3^y-\frac{1}{2}S_4^y-\frac{1}{2}S_5^y-\frac{1}{2}S_6^y\notag\\
    &=&\sin{(\phi+\bm{\nabla}\phi.\textbf{r}_1)}-\sin{(\phi-\bm{\nabla}\phi.\textbf{r}_1)}\notag\\
    &&-\frac{1}{2}(\sin{(\phi+\bm{\nabla}\phi.\textbf{r}_2)}-\sin{(\phi-\bm{\nabla}\phi.\textbf{r}_2)}+\sin{(\phi+\bm{\nabla}\phi.\textbf{r}_3)}
     -\sin{(\phi-\bm{\nabla}\phi.\textbf{r}_3)})\notag\\
     &=& 2\cos{\phi}\sin{(\bm{\nabla}\phi.\textbf{r}_1)}-\frac{1}{2}(2\cos{\phi}\sin{(\bm{\nabla}\phi.\textbf{r}_2)}+2\cos{\phi}\sin{(\bm{\nabla}\phi.\textbf{r}_3)})\notag\\
     &=& \cos{\phi}(2\sin{(\bm{\nabla}\phi.\textbf{r}_1)}-\sin{(\bm{\nabla}\phi.\textbf{r}_2)}-\sin{(\bm{\nabla}\phi.\textbf{r}_3)})\notag\\
     &=& \sin{\phi}(2\bm{\nabla}\phi.\textbf{r}_1-\bm{\nabla}\phi.\textbf{r}_2-\bm{\nabla}\phi.\textbf{r}_3+\mathcal{O}(3))\notag\\
     &\approx& \cos{\phi}(2\textbf{r}_1-\textbf{r}_2-\textbf{r}_3).\bm{\nabla}\phi\notag\\
     &\approx& \cos{\phi}(2\frac{\partial \phi}{\partial x}+\frac{1}{2}\frac{\partial \phi}{\partial x}+\frac{\sqrt{3}}{2}\frac{\partial \phi}{\partial y}+\frac{1}{2}\frac{\partial \phi}{\partial x}-\frac{\sqrt{3}}{2}\frac{\partial \phi}{\partial y})\notag\\
     &\approx& 3\cos{\phi}\frac{\partial \phi}{\partial x} 
     \label{eq:M_33_minus1}
\end{eqnarray}

Now consider
\begin{eqnarray}
    &=&2x_1y_1S_1^x+2x_2y_2S_2^x+2x_3y_3S_3^x+2x_4y_4S_4^x+2x_5y_5S_5^x+2x_6y_6S_6^x\notag\\
    &=&0.S_1^x+0.S_2^x+\frac{\sqrt{3}}{2}(S_3^x+S_4^x)-\frac{\sqrt{3}}{2}(S_5^x+S_6^x)\notag\\
    &=&\frac{\sqrt{3}}{2}(\cos{(\phi+\bm{\nabla}\phi.\textbf{r}_2)}-\cos{(\phi-\bm{\nabla}\phi.\textbf{r}_2)}-\frac{\sqrt{3}}{2}(\cos{(\phi+\bm{\nabla}\phi.\textbf{r}_3)}
     -\cos{(\phi-\bm{\nabla}\phi.\textbf{r}_3)})\notag\\
     &=& \frac{\sqrt{3}}{2}(-2\sin{\phi}\sin{(\bm{\nabla}\phi.\textbf{r}_2)})-\frac{\sqrt{3}}{2}(-2\sin{\phi}\sin{(\bm{\nabla}\phi.\textbf{r}_3)})\notag\\
     &=& \sqrt{3}\sin{\phi}(-\sin{(\bm{\nabla}\phi.\textbf{r}_2)}+\sin{(\bm{\nabla}\phi.\textbf{r}_3)})\notag\\
     &=& \sqrt{3}\sin{\phi}(-\bm{\nabla}\phi.\textbf{r}_2+\bm{\nabla}\phi.\textbf{r}_3+\mathcal{O}(3))\notag\\
     &\approx& \sqrt{3}\sin{\phi}(-\textbf{r}_2+\textbf{r}_3).\bm{\nabla}\phi\notag\\
     &\approx& \sqrt{3}\sin{\phi}(\frac{1}{2}\frac{\partial \phi}{\partial x}+\frac{\sqrt{3}}{2}\frac{\partial \phi}{\partial y}-\frac{1}{2}\frac{\partial \phi}{\partial x}+\frac{\sqrt{3}}{2}\frac{\partial \phi}{\partial y})\notag\\
     &\approx& 3\sin{\phi}\frac{\partial \phi}{\partial y}
     \label{eq:M_33_minus2}
\end{eqnarray}

Using \ref{eq:M_33_minus1} and \ref{eq:M_33_minus2} in \ref{eq:M_33_minus} we obtain
\begin{eqnarray}
M_{33}^-&\approx& \cos{\phi}\frac{\partial \phi}{\partial x}+\sin{\phi}\frac{\partial \phi}{\partial y}\\
&\approx & \bf{N}(\phi).\bf{\nabla}\phi
\end{eqnarray}

\section{Octupolar vortex formation} \label{app:oct}
\begin{figure}
\center
\includegraphics[width=0.9\textwidth]{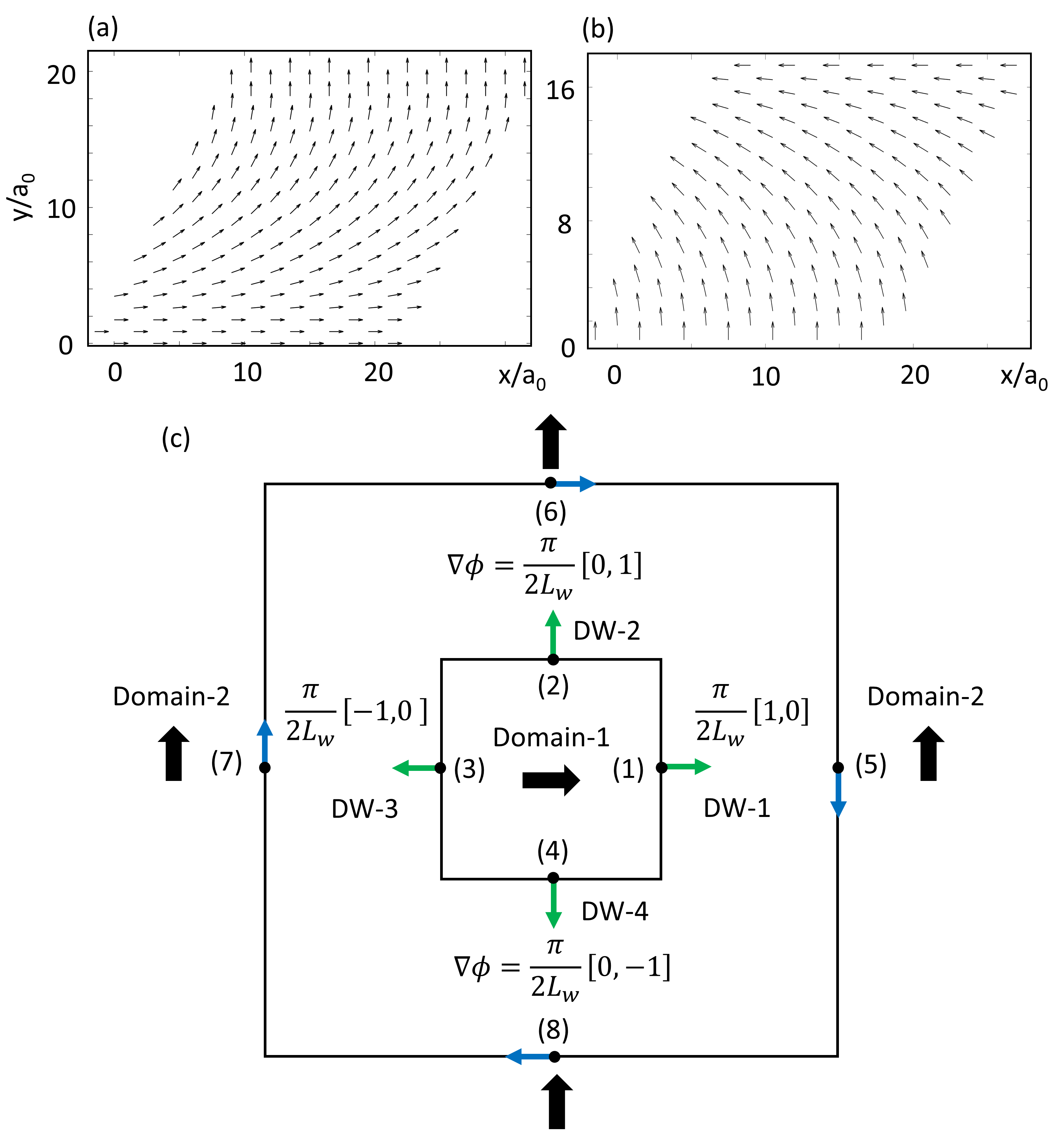} 
\caption{Octupolar moments $M_{33}^+$ and $M_{33}^-$ in uniform $90^\circ$ domain walls. (a) Order parameter of $90^\circ$ uniform domain wall along y-axis. (b) Vector representation of octupolar moments: $\boldsymbol{M}_{33}=[M_{33}^+, M_{33}^-]$ in uniform $90^\circ$ domain wall along y-axis. (c) Vector representation of octupolar moments $M_{33}^+$ and $M_{33}^-$ in domain walls. The black arrows represent the direction of order parameter in two domains and $\boldsymbol{\nabla}\phi$ is the gradient of angle of order parameter within domain walls. The blue and green arrows represents the vector $[{M_{33}^+, M_{33}^-}]$ at $\phi=90^\circ$ and $\phi=0^\circ$ respectively.}
\label{Fig:5}
\end{figure}

To explain the octupole moment vortices, first we consider a vertical domain wall where the order parameter changes uniformly from $0^\circ$ to $90^\circ$ as shown in Fig.~\ref{Fig:5}(a). For this domain wall ${\partial \phi}/{\partial x}=0$,  ${\partial \phi}/{\partial y}=\frac{\pi}{2L_y}$ where $L_y$ is the length of domain wall. The analytical expressions of octupole moments simplify to $M_{33}^-\approx (\frac{\pi}{2L_y})\sin{(\frac{\pi y}{2L_y})}$ and $M_{33}^+\approx (\frac{\pi}{2L_y})\cos{(\frac{\pi y}{2L_y})}$ and and we obtain $\boldsymbol{M}_{33}\approx (\frac{\pi}{2L_y})[\sin{(\frac{\pi y}{2L_y})},\cos{(\frac{\pi y}{2L_y})}]$. This vector is perpendicular to the order parameter and it is consistent with the numerical results of Fig. 4(b) in the main text.

Next we consider a uniform 90$^\circ$ closed domain wall of length $L_w$ that separates two domains as shown in Fig.~\ref{Fig:5}(c). The order parameter is along x-axis in domain-1 and it is along y-axis in domain-2. The angle of order parameter increases in domain-wall-1 along positive x-axis, $(\boldsymbol{\nabla}\phi)_{,1}=[\frac{\pi}{2L_w},0]$, and it decreases in domain-wall-3, $(\boldsymbol{\nabla}\phi)_{,3}=[-\frac{\pi}{2L_w},0]$. Similarly, the angle of order parameter increases in domain-wall-2 along positive y-axis, $(\boldsymbol{\nabla}\phi)_{,2}=[0,\frac{\pi}{2L_w}]$, and it decreases in domain-wall-4, $(\boldsymbol{\nabla}\phi)_{,4}=[0,-\frac{\pi}{2L_w}]$. Using the values of $(\boldsymbol{\nabla}\phi)_{1,2,3,4}$ vectors along the four axes we get:  
\begin{eqnarray}
M_{33,1}^-&\approx& \frac{\pi}{2L_w}\cos{\phi}, \\
M_{33,2}^-&\approx&\frac{\pi}{2L_w}\sin{\phi}, \\
M_{33,3}^-&\approx& -\frac{\pi}{2L_w}\cos{\phi} , \\
M_{33,4}^-&\approx& -\frac{\pi}{2L_w}\sin{\phi} 
\end{eqnarray}
and
\begin{eqnarray}
M_{33,1}^+&\approx&-\frac{\pi}{2L_w}\sin{\phi}, \\
M_{33,2}^+&\approx&\frac{\pi}{2L_w}\cos{\phi}, \\
M_{33,3}^+&\approx& \frac{\pi}{2L_w}\sin{\phi}, \\
M_{33,4}^+&\approx&-\frac{\pi}{2L_w}\cos{\phi}.
\end{eqnarray}
At points (1), (2), (3) and (4), $\phi=0^\circ$ and this lead to $\boldsymbol{M}_{33,1}=\frac{\pi}{2L_w}[1,0]$, $\boldsymbol{M}_{33,2}=\frac{\pi}{2L_w}[0,1]$, $\boldsymbol{M}_{33,3}=\frac{\pi}{2L_w}[-1,0]$ and $\boldsymbol{M}_{33,4}=\frac{\pi}{2L_w}[0,-1]$. These vectors are represented by green arrows as shown in Fig.~\ref{Fig:5}(c) and it rotate around the domain wall. Similarly, for (5), (6), (7) and (8) points, $\phi=90^\circ$ and this lead to $\boldsymbol{M}_{33,5}=\frac{\pi}{2L_w}[0,-1]$, $\boldsymbol{M}_{33,6}=\frac{\pi}{2L_w}[1,0]$, $\boldsymbol{M}_{33,7}=\frac{\pi}{2L_w}[0,1]$ and $\boldsymbol{M}_{33,8}=\frac{\pi}{2L_w}[-1,0]$. These vectors are represented by blue arrows in Fig.~\ref{Fig:5}(c) and it also rotate around the domain wall. As the angle of order parameter varies from $0^\circ$ to $90^\circ$ the vector $\boldsymbol{M}_{33}$ forms vortex like structure and the magnitude of $\boldsymbol{M}_{33}$ is constant.

\section{Representation of $a[V^2]V$ in point group $6/mmm$ ($D_{6h}$)}

As discussed in the main text, the cartesian tensor with Jahn symbol $a[V^2]V$ have 18 free parameters, and can be expanded in terms of the irreps of the point group $6/mmm$ as 
\begin{equation}
    \Gamma^{a[V^2]V}= A_{1u}^-+3 A_{2u}^-+B_{1u}^-+B_{2u}^-+4E_{1u}^-+2E_{2u}^- . 
\end{equation}
All of the irreps that appear in this expansion are both time reversal and spatial inversion odd, as can be seen from the `$u$' subscripts and `$-$' superscripts. As a result, only orders that break both symmetries can give rise to terms in $a[V^2]V$ tensors, such as the spontaneous gyrotropic birefringence. 

The spontaneous gyrotropic birefringence tensor ($\gamma_{ijk}$)is defined as the linear relationship between the $n$'th cartesian component of the wavevector of light ($k_n$) traveling through a medium, and the change in the dielectric impermeability of the medium ($\Delta\beta_{lm}$) as
\begin{equation}
    \Delta \beta_{lm}= i \gamma_{lmn} k_n .
\end{equation}
The magnetotoroidial octupolar components discussed in the main text transform as the irrep $E_{2u}^-$. This irrep appears twice in the expansion of $a[V^2]V$, and hence there are two sets of components of $\gamma$ that transform as the two components $(a,b)$ of $E_u^-(a,b)$. These components are: 
\begin{equation}
    \gamma^I_{E_u^-}=
    \begin{pmatrix}
        0 & 0 & 0\\
        0 & 0 & 0\\
        0 & 0 & 0\\
        a & -b & 0\\
        b & a & 0\\
        0 & 0 & 0
    \end{pmatrix},
\end{equation}
and 
\begin{equation}
    \gamma^{II}_{E_u^-}=
    \begin{pmatrix}
        0 & 0 & b\\
        0 & 0 & -b\\
        0 & 0 & 0\\
        0 & 0 & 0\\
        0 & 0 & 0\\
        0 & 0 & a
    \end{pmatrix}.
\end{equation}
In other words, one can form two separate vectors $\vec{V}^I_{E_u^-}$ and $\vec{V}^{II}_{E_u^-}$ that transform as $E_{2u}^-$ using the components of the $\gamma$ tensor as
\begin{equation}
    \vec{V}^I_{E_u^-}=\left(\gamma_{132}+\gamma_{312}, \gamma_{131}-\gamma_{232}\right)
\end{equation}
and
\begin{equation}
    \vec{V}^{II}_{E_u^-}=\left(2\gamma_{123}, \gamma_{113}-\gamma_{223}\right) .
\end{equation}
